\documentstyle[pre,aps,epic,eepic,epsfig,rotate,preprint]{revtex} 
\begin{document}                                     
\author{Dirk Helbing}
\address{II. Institute of Theoretical Physics, University of Stuttgart,
Pfaffenwaldring 57/III, 70550 Stuttgart, Germany}
\title{Empirical traffic data and their implications for traffic modeling}
\maketitle      
\begin{abstract}
From single vehicle data a number of new empirical results 
about the temporal evolution, correlation, and density-dependence of 
macroscopic traffic quantities have been determined. These have 
relevant implications for traffic modeling and allow to test
existing traffic models.\\[4mm]
\end{abstract}
\pagestyle{myheadings}
\markboth{D. Helbing: Empirical traffic data and their implications for
  traffic modeling, PRE}
{D. Helbing: Empirical traffic data and their implications for traffic 
modeling, PRE}
With the aim of optimizing traffic flow and improving today's
traffic situation, several models for freeway traffic
have been proposed, microscopic \cite{Nagel,Bando,Naga} and macroscopic
ones \cite{LW,Rich,Payne,Kuehne,KK,Hill,Hel,Phil,Nav,MS,PhysA}. 
Only some of them have been systematically derived from
the underlying laws of individual vehicle dynamics \cite{Phil,Nav,MS,PhysA,Wag,Buch}.
Most models are phenomenological in nature \cite{LW,Rich,Payne,Kuehne,KK,Hill}.
These base on various assumptions, the correctness of which has not
been carefully discussed up to now, mainly due to a lack of empirical data
or difficulties to obtain them.
Therefore, this paper presents some fundamental empirical observations
which allow to test some of the models.
\par
The empirical relations have been evaluated from single vehicle data
of both lanes of the Dutch freeway A9 between Haarlem and Amsterdam
(cf. Fig.~\ref{F1}) \cite{Note1}. These data were detected by induction loops
at discrete places $x$ below the lanes $i$ of the roadway and include the 
passage times $t_\alpha(x,i)$,
velocities $v_\alpha(x,i)$ and lengths $l_\alpha(x,i)$ 
of the single vehicles $\alpha$. 
Consequently, it was possible to calculate the number $N_i(x,t)$
of vehicles on lane $i$ which passed the cross section at place $x$
during a time interval $[t,t+T]$, the {\em traffic flow}
\begin{equation}
 {\cal Q}_i(x,t) := N_i(x,t) / T \, ,
\end{equation}
and the macroscopic {\em velocity moments}
\begin{equation}
 \langle v^k \rangle_i := \frac{1}{N_i(x,t)}
 \sum_{t \le t_\alpha < t + T} [v_\alpha(x,i)]^k \, .
\end{equation}
If nothing else is mentioned, the interval length was chosen 
$T = 5$\,min, since this allowed 
to separate the systematic temporal evolution of the macroscopic 
traffic quantities from their statistical fluctuations \cite{Buch}. 
The {\em vehicle densities} $\rho_i(x,t)$ were calculated via the flow formula
\begin{equation}
 {\cal Q}_i(x,t) = \rho_i(x,t) V_i(x,t) \, .
\end{equation}
A detailled comparison with other available methods for the determination
of the average velocities $V_i := \langle v \rangle_i$ and the
vehicle densities $\rho_i$ from single vehicle data will be given in 
\cite{Buch}.
\par
Finally, the {\em lane averages} of the above quantities were defined
according to
\begin{equation}
 {\cal Q}(x,t) := \frac{1}{I} \frac{\sum_i N_i(x,t)}{T}
 = \frac{1}{I} \sum_i {\cal Q}_i(x,t) \, ,
\end{equation}
\begin{equation}
 \langle v^k \rangle := \frac{1}{\sum_j N_j(x,t)}
 \sum_i \sum_{t \le t_\alpha < t + T} [v_\alpha(x,i)]^k \, ,
\end{equation}
and
\begin{equation}
 \rho(x,t) := {\cal Q}(x,t) / V(x,t) \, ,
\end{equation}
where $I$ denotes the number of lanes and
$V(x,t) := \langle v \rangle$. 
Therefore, we have the following relation:
\begin{equation}
 \langle v^k \rangle  = \sum_i 
  \frac{N_i(x,t)}{\sum_j N_j(x,t)} \langle v^k \rangle_i \, .
\end{equation}
\par 
For reasons of simplicity, most macroscopic traffic models
describe the dynamics of the total cross section of the road in an overall 
manner by equations for the density $\rho$ and the average velocity $V$. 
However, one would expect that a realistic description requires a
model of the traffic dynamics on the single lanes and their mutual
coupling due to overtaking and lane-chaning maneuvers \cite{MS,Buch}. This 
could cause a more complex dynamics like density oscillations
among the lanes \cite{osc}. 
\par
\begin{figure}[htbp]
\unitlength8mm
\begin{center}
\begin{picture}(16,10.6)(0.6,-0.8)
\put(0,9.8){\epsfig{height=16\unitlength, width=9.8\unitlength, angle=-90, 
      bbllx=50pt, bblly=50pt, bburx=554pt, bbury=770pt, 
      file=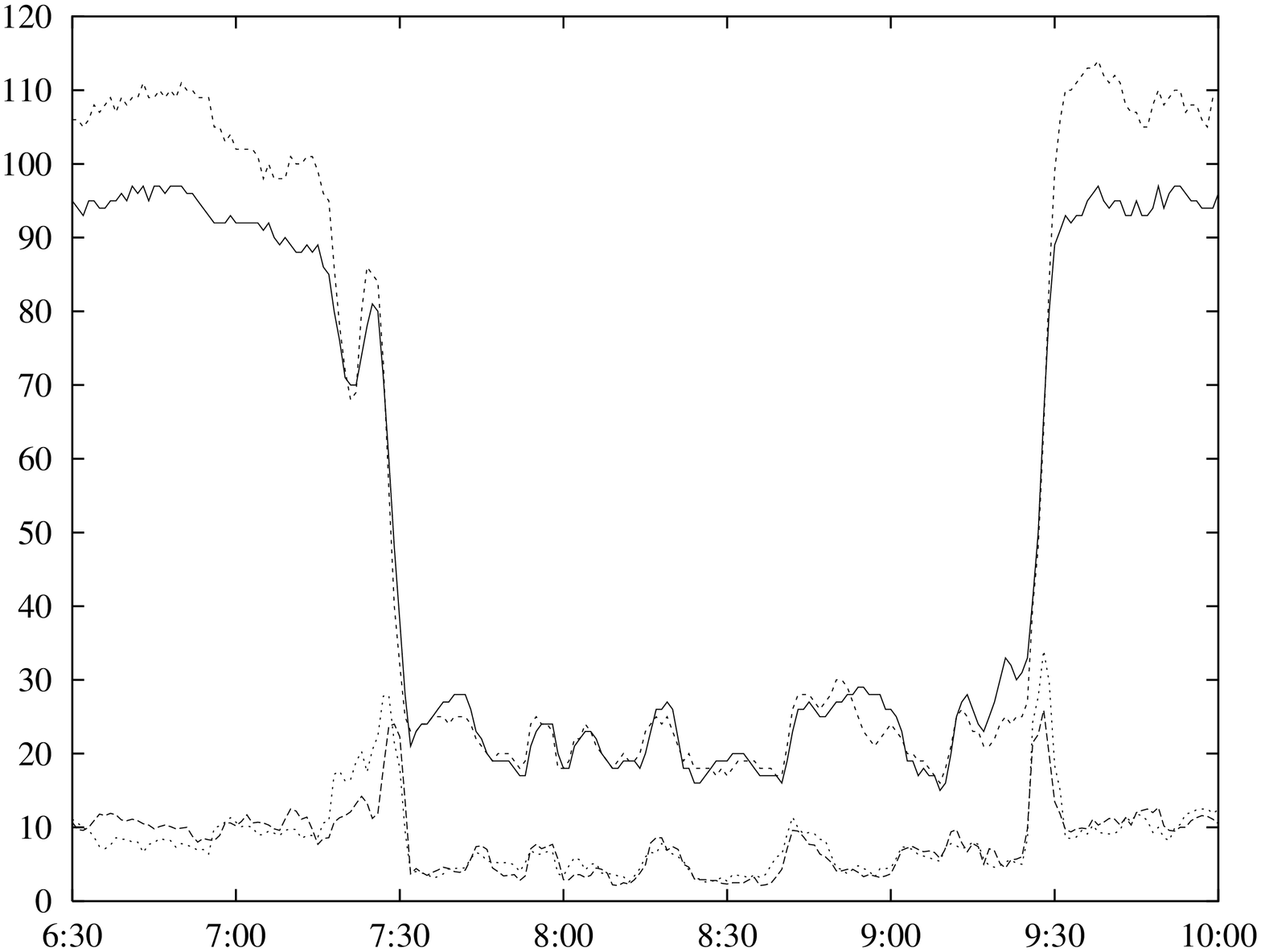}}
\put(8.7,-0.6){\makebox(0,0){\small $t$ (h)}}
\put(8.7,2.9){\makebox(0,0){\small $V_i(x,t)$ (km/h)}}
\put(8.7,1.4){\makebox(0,0){\small $\sqrt{\theta_i}(x,t)$ (km/h)}}
\end{picture}
\end{center}
\caption[]{The temporal course of the average velocities
$V_i$ (---: right lane; -~-~-: left lane) and the standard deviations
$\sqrt{\theta_i}$ of vehicle velocities
(--~--: right lane; $\cdots$: left lane) on November 2, 1994 at
$x=41.75$\,km.}
\label{F3}
\end{figure}

In order to check this, we will
investigate the correlation between neighboring lanes.
Figure \ref{F2} shows that the temporal course of the
densities $\rho_1(x,t)$ and $\rho_2(x,t)$ is almost parallel. A similar thing holds 
for the average velocities $V_1(x,t)$ and $V_2(x,t)$ \cite{Note2}. 
The difference between the curves
is mainly a function of density (cf. Figure \ref{F2a}): At small densities, the vehicles
can move faster on the left lane than on the right one, whereas
at high densities the left lane is more crowded than the right one.
In addition, Figure \ref{F3} shows that the variances
\begin{equation}
 \theta_i(x,t) := \langle (v - V_i )^2 \rangle_i 
 = \langle v^2 \rangle_i - (\langle v \rangle_i)^2
\end{equation}
behave almost identically on the neighboring lanes (although
the order of magnitude of the average velocities $V_i(x,t)$ 
changes considerably). This strong correlation between neighboring
lanes probably arises from overtaking and lane-changing maneuvers.
It justifies the common practice to describe the dynamics of the total
cross section of the road in an overall manner.
\par
\begin{figure}[htbp]
\unitlength8mm
\begin{center}
\begin{picture}(16.2,10.6)(-0.2,-0.8)
\put(0,9.8){\epsfig{height=16\unitlength, width=9.8\unitlength, angle=-90, 
      bbllx=50pt, bblly=50pt, bburx=554pt, bbury=770pt, 
      file=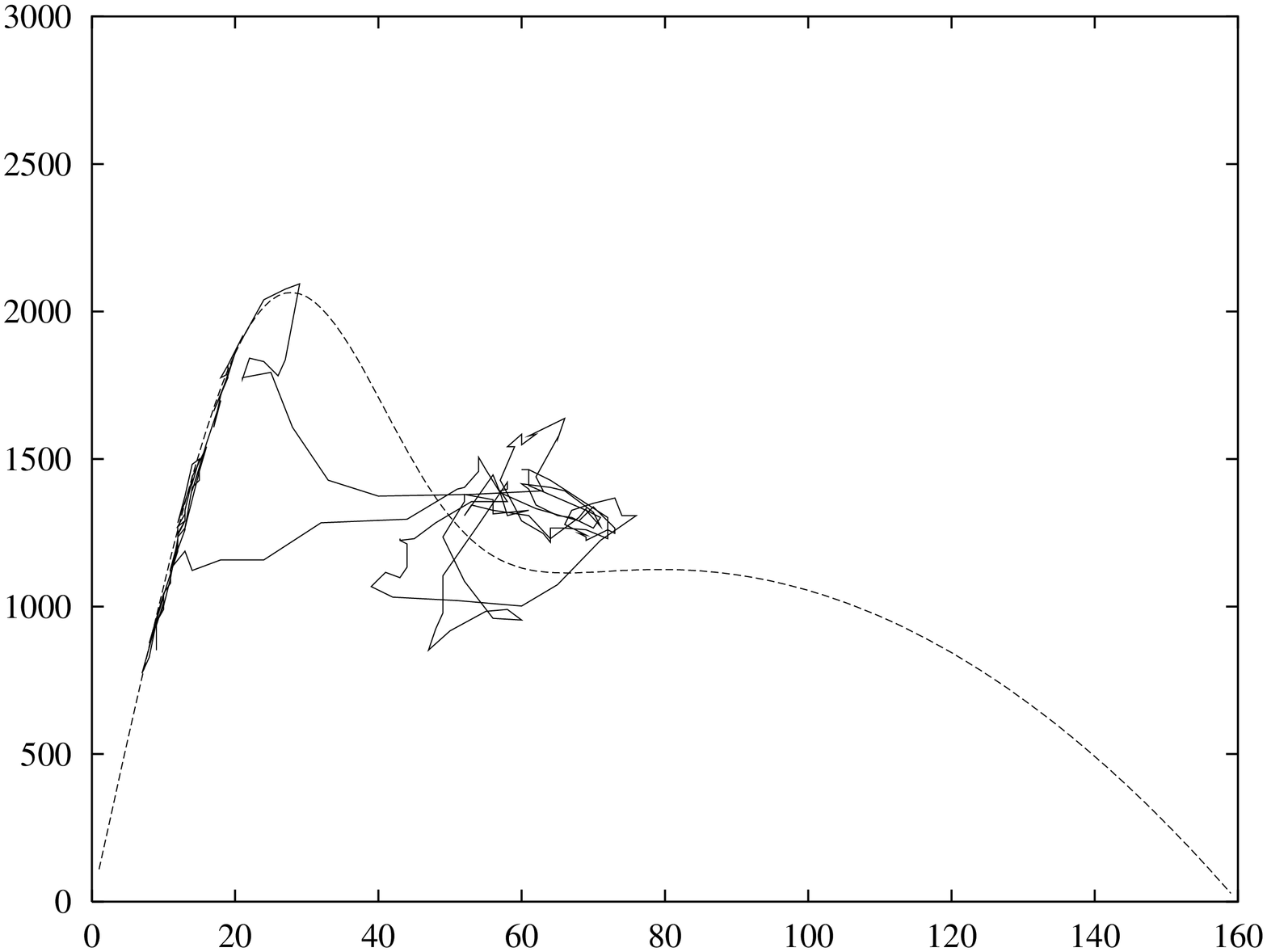}}
\put(8.7,-0.6){\makebox(0,0){\small $\rho$ (vehicles/km)}}
\put(0,5.1){\makebox(0,0){\rotate[l]{\hbox{\small ${\cal Q}$ (vehicles/h)}}}}
\thicklines
\put(4.12,5.75){\vector(1,-2){0.15}}
\put(3.8,4.005){\vector(-1,0){0.2}}
\thinlines
\end{picture}
\end{center}
\caption[]{Comparison of the {\em fundamental diagram} ${\cal Q}_e(\rho)$
(i.e. the average flow-density relation, --~--) 
with the temporal evolution of traffic flow ${\cal Q}(x,t)
= \rho(x,t) V(x,t)$ (---) at $x=41.75$\,km.}
\label{F4}
\end{figure}

Now we face the question, how a realistic traffic model must look like.
Due to the conservation of the number of vehicles, the dynamics of the
vehicle density is given by the {\em continuity equation} 
\cite{LW,Rich,Payne,MS,Buch}
\begin{equation}
 \frac{\partial \rho(x,t)}{\partial t} + \frac{\partial {\cal Q}(x,t)}{\partial x}
 = \nu^+(x,t) - \nu^-(x,t) \, ,
\end{equation}
where $\nu^+(x,t)$ and $\nu^-(x,t)$ are the rates of vehicles which enter or leave
the freeway at on- and off-ramps, respectively. Lighthill, Whitham and
Richards have suggested to specify the flow ${\cal Q}(x,t)$ in accordance with
an empirical flow-density relation ${\cal Q}_e(\rho)$ \cite{LW,Rich}:
\begin{equation}
 {\cal Q}(x,t) = {\cal Q}_e(\rho(x,t)) \, .
\label{equi}
\end{equation}
This relation has been called into question, since the resulting model
cannot describe the emergence of phantom traffic jams or
stop-and-go traffic \cite{Hel,Buch}. Therefore,
some researchers have introduced an additional dynamical equation for the average 
velocity $V(x,t)$ which allows to describe instabilities of traffic
flow \cite{Kuehne,KK,Hill,Hel,PhysA}. However, others
have interpreted these phenomena as effects of fluctuations or of
phantom bottlenecks caused by slow, overtaking vehicles like trucks \cite{Phantom}.
Hence we check relation (\ref{equi}) in Figure \ref{F4}. 
It is found that (\ref{equi}) becomes invalid above a density of
about 12 vehicles per kilometer and lane, where a hysteresis effect
occurs \cite{Note3}. This indicates a transition from stable to unstable
traffic flow. 
\par
\begin{figure}[htbp]
\unitlength8mm
\begin{center}
\begin{picture}(16,10.6)(0,-0.8)
\put(0,9.8){\epsfig{height=16\unitlength, width=9.8\unitlength, angle=-90, 
      bbllx=50pt, bblly=50pt, bburx=554pt, bbury=770pt, 
      file=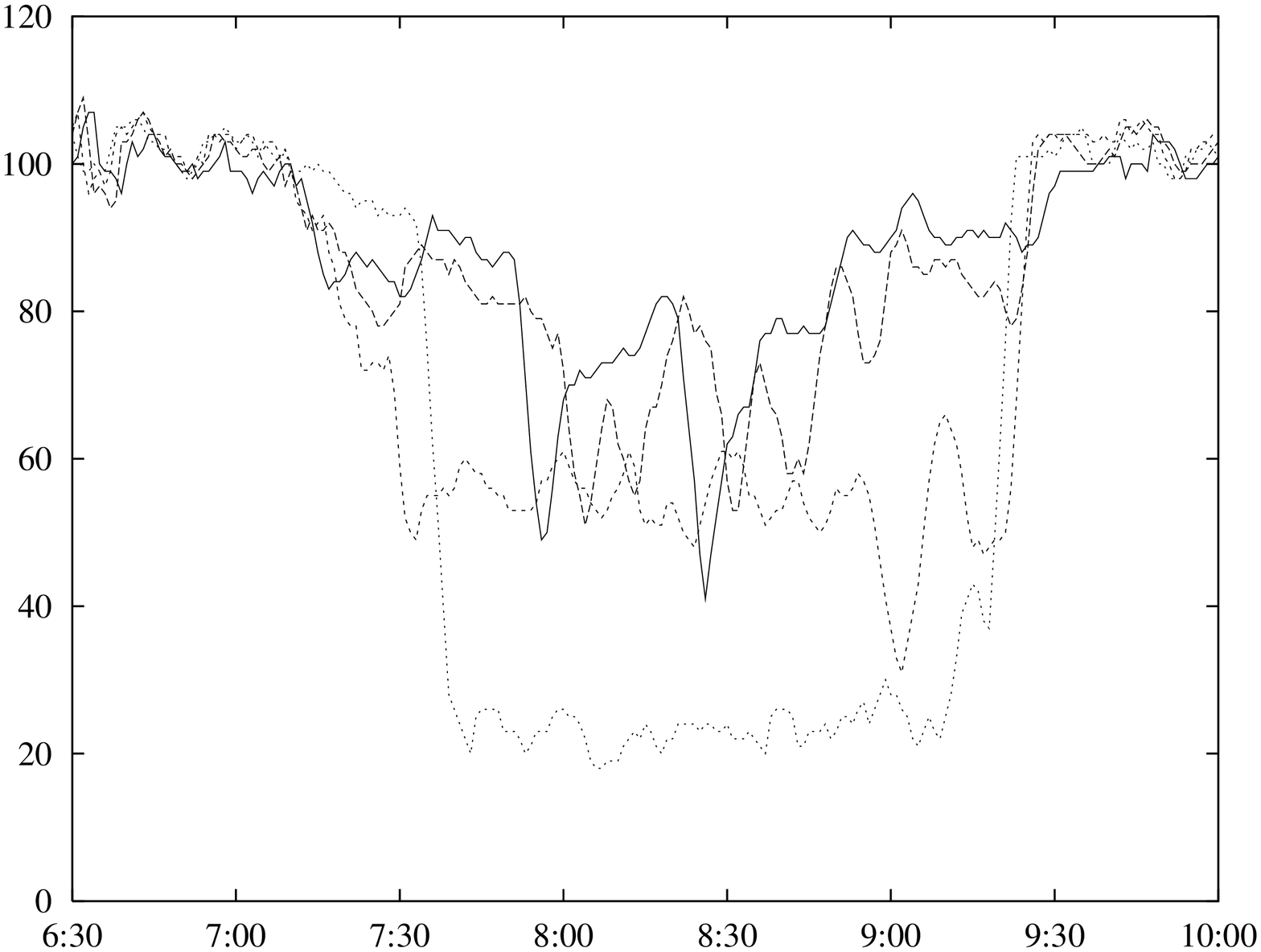}}
\put(8.7,-0.6){\makebox(0,0){\small $t$ (h)}}
\put(0.2,5.1){\makebox(0,0){\rotate[l]{\hbox{\small $V(x,t)$ (km/h)}}}}
\end{picture}
\end{center}
\caption[]{Temporal evolution of the average velocity
$V(x,t)$ at successive cross sections of the freeway on
October 14, 1994 ($\cdots$: $x=41.75$\,km; -~-~-: $x=40.8$\,km;
--~--: $x=39.6$\,km; ---: $x=37.6$\,km).}
\label{F5}
\end{figure}

An empirical proof of emerging stop-and-go traffic is presented in
Figure \ref{F5}. During the rush hours between 7:30\,am and 9:30\,am, 
average velocity breaks down at place $x=41.75$\,km because of the
on-ramp at $x=41.3$\,km. Nevertheless, the traffic situation recovers
at the successive cross sections, i.e. average
velocity increases again. In spite of this, the initially small 
velocity oscillations at $x = 41.75$\,km grow considerably
in the course of the road. This corresponds to emerging stop-and-go 
traffic (i.e. alternating periods of acceleration and deceleration).
At the same time, the wavelength of the oscillation increases. This is in
good agreement with computer simulations which show a merging of
density clusters leading to larger wave lengths \cite{KK}. 
\par
\begin{figure}[htbp]
\unitlength8mm
\begin{center}
\begin{picture}(16,10.6)(0,-0.8)
\put(0,9.8){\epsfig{height=16\unitlength, width=9.8\unitlength, angle=-90, 
      bbllx=50pt, bblly=50pt, bburx=554pt, bbury=770pt, 
      file=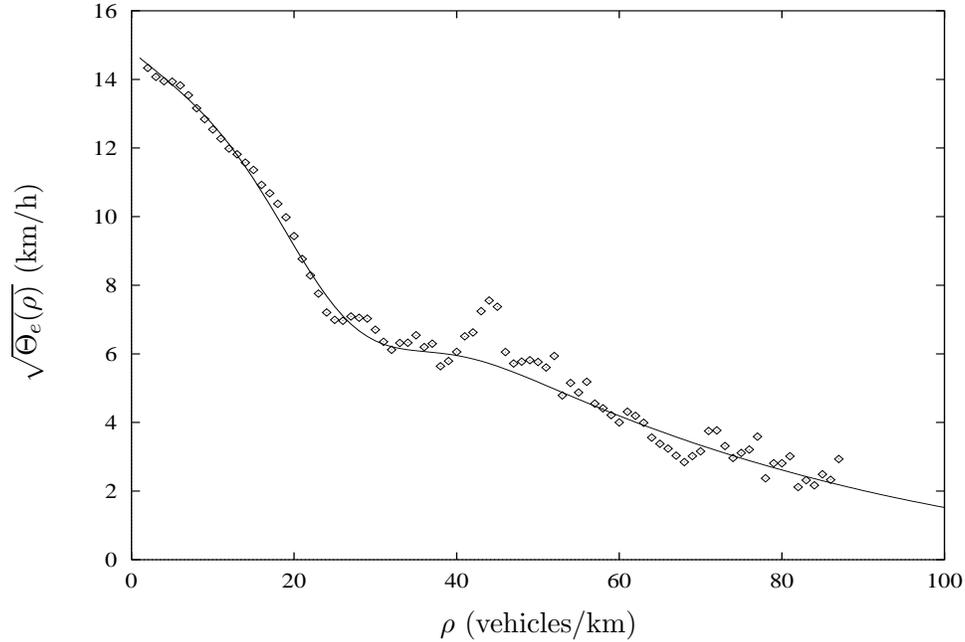}}
\put(8.7,-0.6){\makebox(0,0){\small $\rho$ (vehicles/km)}}
\put(0.2,5.1){\makebox(0,0){\rotate[l]{\hbox{\small
        $\sqrt{\Theta_e(\rho)}$ (km/h)}}}}
\end{picture}
\end{center}
\caption[]{Average density-dependence of the
standard deviation $\sqrt{\Theta_e}$ of vehicle velocities
(for $T=1$\,min, $\Diamond$) and corresponding fit function (---).}
\label{F6a}
\end{figure}

\begin{figure}[htbp]
\unitlength8mm
\begin{center}
\begin{picture}(16,10.6)(0.6,-0.8)
\put(0,9.8){\epsfig{height=16\unitlength, width=9.8\unitlength, angle=-90, 
      bbllx=50pt, bblly=50pt, bburx=554pt, bbury=770pt, 
      file=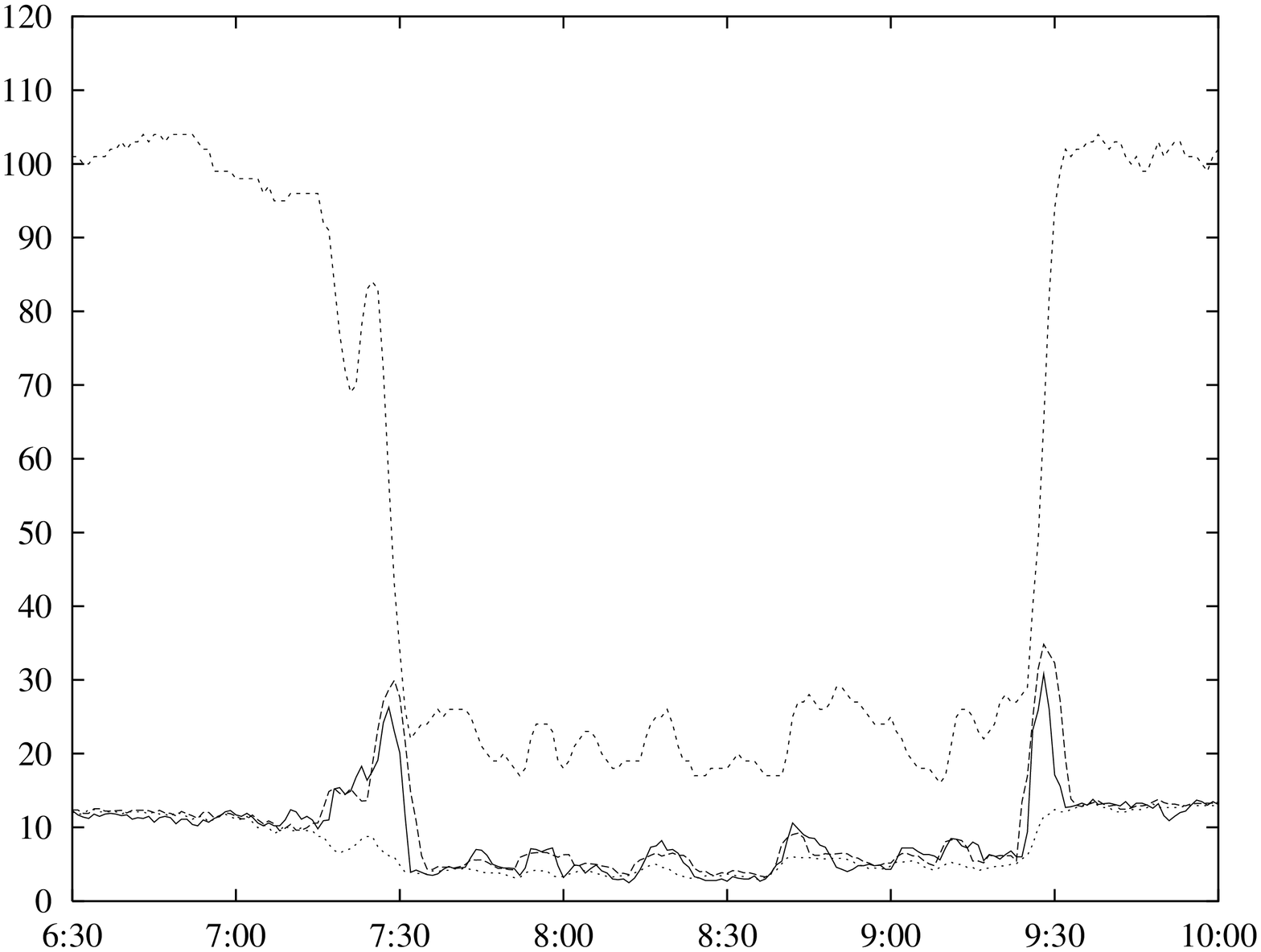}}
\put(8.7,-0.6){\makebox(0,0){\small $t$ (h)}}
\put(8.7,2.7){\makebox(0,0){\small $V(x,t)$ (km/h)}}
\put(8.7,1.4){\makebox(0,0){\small $\sqrt{\Theta}(x,t)$ (km/h)}}
\end{picture}
\end{center}
\caption[]{Temporal evolution of the standard deviation
$\sqrt{\Theta(x,t)}$ of vehicle velocities (---) in comparison with
the equilibrium approximation $\sqrt{\Theta_e(\rho(x,t))}$ ($\cdots$).
Obviously, the empirical variance $\Theta(x,t)$ has peaks
where the average velocity $V(x,t)$ (-~-~-) changes considerably.
In order to describe these, a correction term due to time averaging
must be added (--~--). (Data from November 2, 1994 at $x=41.75$\,km)}
\label{F6}
\end{figure}

 
After we have found that we need a dynamic velocity equation for an adequate
description of the spatio-temporal evolution of
traffic flow, we have to clear up the question, whether
we also need a dynamic equation for the variance 
\begin{equation}
 \Theta := \langle (v - V)^2 \rangle 
 = \theta + \langle (V_i - V)^2 \rangle 
\end{equation}
or not. Theoretical considerations on the basis of gas-kinetic approaches 
have shown that the velocity equation depends on the variance, for 
which a separate equation can be derived \cite{Phil,Nav,MS,PhysA,Wag,Buch}. Nevertheless we
will try out the equilibrium approximation
\begin{equation}
 \Theta(x,t) = \Theta_e(\rho(x,t)) \, ,
\end{equation}
where $\Theta_e(\rho)$ is the empirical variance-density relation 
(cf. Figure \ref{F6a}). Figure \ref{F6} shows that this approximation fits
the temporal evolution of the variance in a satisfactory way 
as long as the average
velocity $V$ does not rapidly change. However, when the velocity breaks
down or increases, the variance shows mysterious peaks. These are
a consequence of having built the temporal averages 
\begin{equation}
 \langle v^k \rangle (x,t) \equiv \overline{\langle v^k \rangle_a}(x,t)
 := \frac{1}{T} \!\!\int_t^{t+T}\!\!
 dt' \,  \langle v^k \rangle_a(x,t')
\end{equation}
over finite time intervals $T$, where
$\langle v^k \rangle_a(x,t) := \int dv \, v^k P_a(v;x,t)$ with
the actual velocity distribution $P_a(v;x,t)$. 
In linear Taylor approximation we find
\begin{eqnarray}
 V(x,t) 
 &\approx & \frac{1}{T} \!\!\int_t^{t+T}\!\!
 dt' \, \Big[ V_a(x,t) + \frac{\partial V_a(x,t)}{\partial t}
 (t' - t) \Big] \nonumber \\  
 &=& V_a(x,t) + \frac{T}{2} \frac{\partial
   V_a(x,t)}{\partial t} \, .
\end{eqnarray}
Since $\partial V_a/\partial t$ is varying around zero,
the measured value $V$ fluctuates around the actual value 
$V_a(x,t) := \langle v \rangle_a(x,t)$. For the variance we find
\begin{eqnarray}
 \Theta &\equiv & \overline{\langle ( v - 
 V )^2 \rangle_a }
 = \overline{\Theta_a} + \overline{[V_a - 
 V ]^2}  \nonumber \\
 &=& \overline{\Theta_a} + \frac{T^2}{4}
 \overline{\left( \frac{\partial V_a}{\partial t} \right)^2} \, .
\end{eqnarray}
Therefore, time averaging leads to a positive correction term which 
becomes particulary large, where the average velocity changes rapidly,
but vanishes in the limit $T \rightarrow 0$. 
This correction term describes
the variance peaks in Figure \ref{F6} quite well. Consequently, 
the dynamics of the variance can be reconstructed from the
dynamics of the vehicle density $\rho(x,t)$ and the average velocity
$V(x,t)$.
\par
Summarizing our results, we were able to demonstrate the following
by empirical data:
1. The dynamics of neighboring lanes is strongly correlated so that
the total freeway cross section can be described in an overall way.
2. There is a transition from stable to unstable traffic flow
at a critical density $\rho_{cr}$ of about 12 vehicles per kilometer and lane.
3. Emergent stop-and-go traffic exists, so that a realistic traffic
model must contain a dynamic velocity equation.
4. The variance can be well approximated by an equilibrium relation, if
corrections due to time averaging are taken into account. 
These conclusions seem to be also valid for other stretches of 
freeway systems, at least European ones. 
\par
The empirical findings
question the fluid-dynamic model by Lighthill, Whitham and Richards
\cite{LW,Rich}. They are in favour of the phenomenological models 
by Payne \cite{Payne}, Phillips \cite{Phil}, K\"uhne \cite{Kuehne},
Kerner and Konh\"auser \cite{KK}, Hilliges \cite{Hill} as well as a recent
model by Helbing \cite{PhysA,Buch} which has been systematically derived from the microscopic
vehicle dynamics via a gas-kinetic level of description. 
The last of these models fits the instability region best, 
in particular the surprisingly low critical density $\rho_{cr}$  
\cite{PhysA,Buch}.

\section*{Acknowledgments}

The author is grateful to Henk Taale and the {\it Ministry of Transport,
Public Works and Water Management} for supplying the freeway data.

\clearpage
\begin{figure}[htbp]
\unitlength1.6cm
\begin{center}
\begin{picture}(9.42,2.2)
\thicklines
\put(1.53,1.8){\makebox(0,0){\small Rottepolderplein}}
\put(2.83,1.8){\makebox(0,0){\small S\,17\vphantom{p}}}
\put(7.56,1.8){\makebox(0,0){\small Badhoevedorp}}
\put(4.51,1.3){\vector(1,0){0.4}}
\put(4.51,1.1){\vector(1,0){0.4}}
\put(0.3,1.3){\makebox(0,0){\small $i=2$}}
\put(0.3,1.1){\makebox(0,0){\small $i=1$}}
\thinlines
\put(0,1.4){\line(1,0){9.42}}
\dashline{0.1}(0,1.2)(9.42,1.2)
\put(1,0.5){\line(0,1){0.9}}
\put(2.06,0.5){\line(0,1){0.9}}
\put(2.56,0.5){\line(0,1){0.9}}
\dashline{0.07}(3.01,0.5)(3.01,1.4)
\put(3.51,0.5){\line(0,1){0.9}}
\put(4.71,0.5){\line(0,1){0.9}}
\put(6.71,0.5){\line(0,1){0.9}}
\put(7.41,0.5){\line(0,1){0.9}}
\put(7.71,0.5){\line(0,1){0.9}}
\put(8.42,0.5){\line(0,1){0.9}}
\put(1,0.2){\makebox(0,0){\small 43.31\,km}}
\put(1.96,0.2){\makebox(0,0){\small 42.25}}
\put(2.66,0.2){\makebox(0,0){\small 41.75}}
\put(3.51,0.2){\makebox(0,0){\small 40.80}}
\put(4.71,0.2){\makebox(0,0){\small 39.60}}
\put(6.51,0.2){\makebox(0,0){\small 37.60}}
\put(7.21,0.2){\makebox(0,0){\small 36.90}}
\put(7.91,0.2){\makebox(0,0){\small 36.60}}
\put(8.82,0.2){\makebox(0,0){\small 35.89\,km}}
\put(0,1){\line(1,0){0.75}}
\put(0.75,1){\line(1,-1){0.15}}
\put(1,1){\line(1,-1){0.15}}
\put(1,1){\line(1,0){1.06}}
\put(2.06,1){\line(-1,-1){0.15}}
\put(2.31,1){\line(-1,-1){0.15}}
\put(2.31,1){\line(1,-1){0.15}}
\put(2.56,1){\line(1,-1){0.15}}
\put(2.56,1){\line(1,0){0.45}}
\put(3.01,1){\line(-1,-1){0.15}}
\put(3.26,1){\line(-1,-1){0.15}}
\put(3.26,1){\line(1,0){3.9}}
\put(7.16,1){\line(1,-1){0.15}}
\put(7.41,1){\line(1,-1){0.15}}
\put(7.41,1){\line(1,0){0.3}}
\put(7.71,1){\line(-1,-1){0.15}}
\put(7.96,1){\line(-1,-1){0.15}}
\put(7.96,1){\line(1,0){1.46}}
\end{picture}
\end{center}
\caption[]{The investigated stretch of the Dutch two-lane freeway A9
from Haarlem to Amsterdam including on- and off-ramps. Detectors are
indicated by vertical lines. The detector at $x=41.3$\,km (-~-~-)
only evaluates on-ramp traffic and a bus lane. Between $x=40.8$\,km
and $x=37.6$\,km traffic flow is not disturbed over more than 
three kilometers. The speed limit is 120\,km/h.}
\label{F1}
\end{figure}
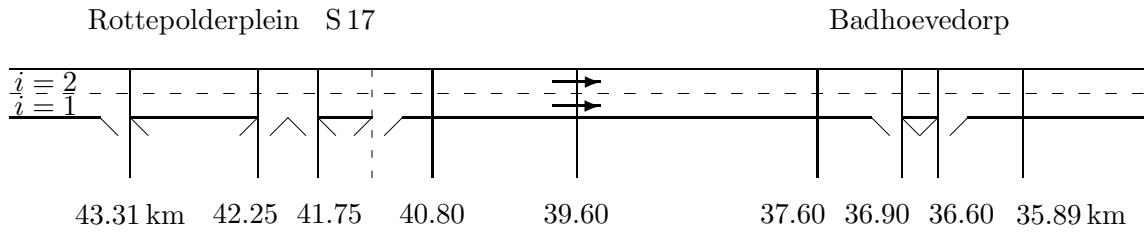
\begin{figure}[htbp]
\unitlength8mm
\begin{center}
\begin{picture}(16,10.6)(0.6,-0.8)
\put(0,9.8){\epsfig{height=16\unitlength, width=9.8\unitlength, angle=-90, 
      bbllx=50pt, bblly=50pt, bburx=554pt, bbury=770pt, 
      file=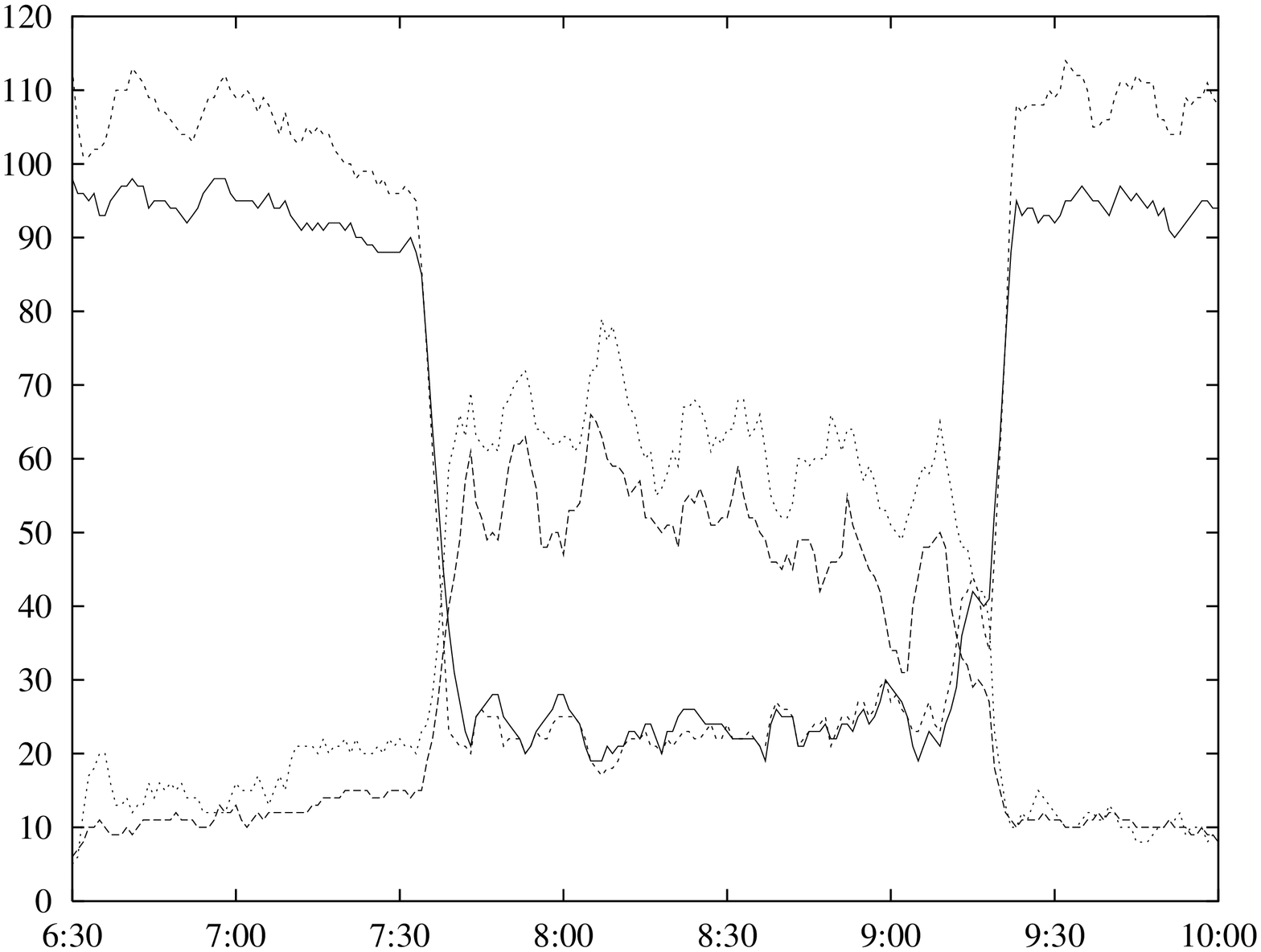}}
\put(8.7,-0.6){\makebox(0,0){\small $t$ (h)}}
\put(8.7,6.8){\makebox(0,0){\small $\rho_i(x,t)$ (vehicles/km)}}
\put(8.7,2.9){\makebox(0,0){\small $V_i(x,t)$ (km/h)}}
\end{picture}
\end{center}
\caption[]{The temporal course of the average velocities
$V_i$ (---: right lane; -~-~-: left lane) and the vehicle densities
$\rho_i$ (--~--: right lane; $\cdots$: left lane) on October 14, 1994
at $x=41.75$\,km.}
\label{F2}
\end{figure}
\begin{figure}[htbp]
\unitlength8mm
\begin{center}
\begin{picture}(16,10.6)(0.6,-0.8)
\put(0,9.8){\epsfig{height=16\unitlength, width=9.8\unitlength, angle=-90, 
      bbllx=50pt, bblly=50pt, bburx=554pt, bbury=770pt, 
      file=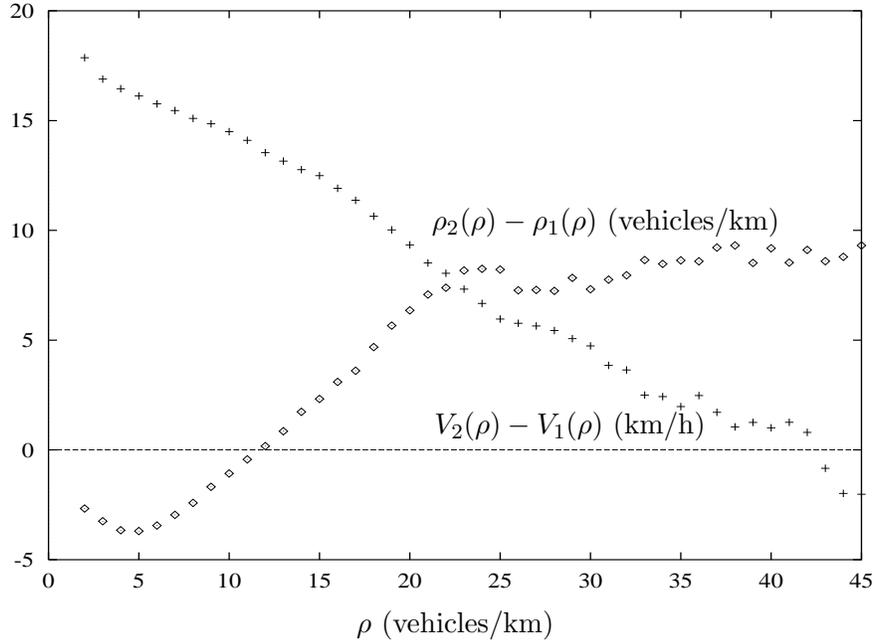}}
\put(8.7,-0.6){\makebox(0,0){\small $\rho$ (vehicles/km)}}
\put(11.2,6.1){\makebox(0,0){\small $\rho_2(\rho) - \rho_1(\rho)$ 
(vehicles/km)}}
\put(10.6,2.7){\makebox(0,0){\small $V_2(\rho) - V_1(\rho)$ (km/h)}}
\end{picture}
\end{center}
\caption[]{Average differences between the vehicle densities
($\Diamond$) and the average velocities (+) on both lanes of the 
Dutch freeway A9 on November 2, 1994 ($T=1$\,min).}
\label{F2a}
\end{figure}

\begin{references}
\bibitem{Nagel} K. Nagel and M. Schreckenberg, J. Physique I
France {\bf 2}, 2221 (1992); K. Nagel and S. Rasmussen, in
{\em Artificial Life IV}, edited by R. A. Brooks and P. Maes
(MIT Press, Cambridge, MA, 1994);
M. Schreckenberg, A. Schadschneider, K. Nagel, and
N. Ito, Phys. Rev. E {\bf 51}, 2939 (1995);
K. Nagel, Phys. Rev. E {\bf 53}, 4655 (1996).
\bibitem{Bando} M. Bando, K. Hasebe, A. Nakayama, A. Shibata,
and Y. Sugiyama, Phys. Rev. E {\bf 51}, 1035 (1995).
\bibitem{Naga} T. Nagatani, J. Phys. A {\bf 28}, L119 (1995);
T. Nagatani, Physica A {\bf 223}, 137 (1996).
\bibitem{LW} M. J. Lighthill and G. B. Whitham, Proc. R. Soc. A {\bf 229},
317 (1955).
\bibitem{Rich} P. I. Richards, Oper. Res. {\bf 4}, 42 (1956).
\bibitem{Payne} H. J. Payne, in {\em Mathematical Models of Public Systems},
edited by G. A. Bekey (Simulation Council, La Jolla, CA, 1971), Vol. 1. 
\bibitem{Kuehne} R. D. K\"uhne, in {\em Proceedings of the 9th International
    Symposium on Transportation and Traffic Theory}, edited by
I. Volmuller and R. Hamerslag (VNU Science, Utrecht, 1984).
\bibitem{KK} B. S. Kerner and P. Konh\"auser, Phys. Rev. E {\bf 48},
R2335 (1993); B. S. Kerner and P. Konh\"auser, Phys. Rev. E {\bf 50},
54 (1994).
\bibitem{Hill} M. Hilliges and W. Weidlich, Transportation Research
{\bf 29}, 407 (1995).
\bibitem{Hel} D. Helbing, Phys. Rev. E {\bf 51}, 3164 (1995).
\bibitem{Phil} W. F. Phillips, Transportation Planning and Technology
{\bf 5}, 131 (1979).
\bibitem{Nav} D. Helbing, Phys. Rev. E {\bf 53}, 2366 (1996).
\bibitem{MS} D. Helbing, in {\em Traffic and Granular Flow}, edited by
D. E. Wolf, M. Schreckenberg, and A. Bachem (World Scientific, Singapore,
1996); D. Helbing and A. Greiner, Modeling and simulation
of multi-lane traffic flow, Phys. Rev. E, submitted (1996).
\bibitem{PhysA} D. Helbing, Derivation and empirical validation of a
refined traffic flow model, Physica A, in print (1996).
\bibitem{Wag} C. Wagner {\it et al.}, Second order continuum traffic
flow model, Phys. Rev. E, submitted (1996).
\bibitem{Buch} D. Helbing, {\em Verkehrsdynamik. Neue physikalische
    Modellierungskonzepte} (Springer, Berlin, in preparation).
\bibitem{Note1} Note that the situation on European freeways
is somewhat different from American ones due to other legal regulations: 
Because of the higher speed limit (if there is any), 
overtaking is only allowed on the left-hand lane. 
Therefore, trucks mainly use the right lane, on which the average
velocity is lower (cf. Fig. \ref{F2a}).
\bibitem{osc} D. C. Gazis, R. Herman, and G. H. Weiss, Oper. Res.
{\bf 10}, 658.
\bibitem{Note2} Note that a
synchronized state of traffic in the sense of B. S. Kerner and 
H. Rehborn [Phys. Rev. E {\bf 53}, R4275] (i.e. with the same velocities
$V_i(x,t)$ on all lanes $i$) only occurs
above an average density $\rho$ of about 35 vehicles per kilometer and lane
(cf. Fig. \ref{F2a}).
\bibitem{Phantom} D. C. Gazis and R. Herman, Transpn. Sci. {\bf 26}, 223
(1992). 
\bibitem{Note3} Similar results have been obtained by J. Treiterer {\it et
    al.} for American freeways 
[in {\em Proceedings of the 6th International Symposium on Transportation and
  Traffic Theory}, edited by D. Buckley (Sydney, 1974)].
\end{references}
\end{document}